\documentclass[%
reprint,
 amsmath,amssymb,
 aps,
]{revtex4-1}
\usepackage{graphicx}% Include figure files
\usepackage{dcolumn}% Align table columns on decimal point
\usepackage{bm}% bold math
\usepackage{url}
\begin{document}

\preprint{APS/123-QED}

\title{Tuning Self Organization Of Confined Active Particles By Steepness
  Of The Trap}
\author{Shubhashis Rana} 
\email{Shubhashis.rana@gmail.com}
\affiliation {S. N. Bose National Centre For Basic Sciences, Kolkata,
  India 700098}
\author{Md.\phantom{x}Samsuzzaman}
\email{samsuzz@gmail.com}
\author{Arnab Saha}
\email{Correspondence to: sahaarn@gmail.com}
\affiliation {Department of Physics, Savitribai Phule Pune University, Pune, 
India 411007}

\date{\today}% It is always \today, today,
             %  but any date may be explicitly specified

\begin{abstract}

We consider collective dynamics of self-propelling particles in two dimensions. They can align themselves according to the direction of propulsion of their neighbours, together with a random perturbation (i.e. rotational fluctuation). They are also interacting with each other by a soft, isotropic, excluded-volume interaction. Particles are confined in a circular trap. The steepness of the trap is tuneable. Their average packing fraction and strength of rotational fluctuation are low. When the trap is steep, particles flock along its boundary. They form polar cluster that spreads over the boundary. The cluster has no spatial or structural order. We show, when the steepness is decreased beyond a threshold value, the clusters become round, compact and eventually spatial order (hexagonal) emerges in addition to the pre-established polar order within them. We investigate kinetics of such ordering. We find that while rotating around the centre of the trap along its circular boundary, the clusters need to roll around their centre of mass, to be spatially ordered. We also discuss the stability of the order when the trap is suddenly switched off.       
  
%\begin{description}
%\item[Usage]
%Secondary publications and information retrieval purposes.
%\item[PACS numbers]
%May be entered using the \verb+\pacs{#1}+ command.
%\item[Structure]
%You may use the \texttt{description} environment to structure your abstract;
%use the optional argument of the \verb+\item+ command to give the category of 
%each item. 
%\end{description}
\end{abstract}

\pacs{Valid PACS appear here}% PACS, the Physics and Astronomy
                             % Classification Scheme.
%Use showkeys class 
					      %option if keyword display desired
\maketitle

%\tableofcontents

%\begin{multicols}{2}

\section{Introduction}

Active (self-propelled) interacting particles are prone to be phase separated to form clusters with varied dynamical and structural properties, even in two dimensions (2D). By now this has been extensively studied in bulk via theory and experiments \cite{deseigne2010collective, cates2010arrested, peruani2012f, schwarz2012phase, palacci2013living, liu2017motility, theurkauff2012dynamic, buttinoni2013dynamical, redner2013structure, cates2013active, fily2012athermal, farrell2012pattern, peruani2011traffic, peruani2010cluster, ginot2018aggregation, peruani2006nonequilibrium, tung2016micro, weitz2015self, stenhammar2014phase, fily2014freezing, levis2014clustering, stenhammar2013continuum, mognetti2013living, patch2017kinetics, martin2018collective, van2019interrupted, sese2018velocity, shi2018self, alarcon2017morphology}.  Passive equilibrated systems lack self-propulsion and self-aligning interactions, particularly of non-steric origin. Hence the dynamics and the patterns observed in the active systems are usually more intricate than the systems in equilibrium. Though surprisingly, it turns out that intriguing physics of two step melting (solid to hexatic to liquid) \cite{halperin1978theory, nelson1979dislocation, strandburg1988two, bernard2011two, engel2013hard, kapfer2015two}, a characteristic feature of 2D passive (equilibrated) systems, may be important to understand the dynamics of clustering and the structures within, observed in self-propelling but not self-aligning soft discs \cite{klamser2018thermodynamic, klamser2018kinetic}. Recent simulations suggest that, similar to equilibrium systems, self-propelling discs, while interacting with each other by soft, isotropic, repulsive potentials can exhibit hexatic phase between solid and liquid phases while melting, even for considerably high activity. According to \cite{klamser2018thermodynamic}, the system needs to be highly active to open up another criticality (independent of the melting one) where self-propelling particles, thanks to their motility, come together to phase separate and form clusters coexisting with gas-like fluid phase. The motility induced phase separation (MIPS) \cite{cates2015motility} is the predominant feature of the active particles where self-propulsion and isotropic excluded volume interactions dominate over self-aligning interactions. It is an outcome of a positive feedback between particle-accumulation and  their motility\cite{stenhammar2013continuum, gonnella2015motility, cates2015motility, wittkowski2014scalar, marenduzzo2016introduction}. Though MIPS is reported frequently in the current literature but its characterisation remains illusive. Various issues related to MIPS, e.g. whether the dense phase obtained in MIPS is liquid-like \cite{klamser2018thermodynamic} or solid-like \cite{bialke2012crystallization, redner2013structure} or polycrystalline \cite{cugliandolo2017phase},  whether the kinetics of phase separation in all the active systems are equivalent or not \cite{redner2013structure, buttinoni2013dynamical, briand2016crystallization} etc., are debatable. 

The other extreme, where aligning interaction dominates over excluded volume interaction in a system of self-propelling particles, with sufficiently low fluctuations and high density, transitions from isotropic to ordered phases are observed, instead of MIPS. Depending on the types of the aligning interactions \cite{vicsek1995novel, chate2006simple, liebchen2017collective}, choice of the parameter space \cite{caussin2014emergent}, underlying symmetry of  the system together with non-equilibrium fluctuations at large scales \cite{solon2013revisiting, solon2015phase, solon2015flocking} various kinds of ordered phases (e.g. - nematic, polar, smectic) and ordered structures (e.g. - periodic density waves, solitonic bands, polar-liquid droplet, rotating micro-flocks etc.) \cite{solon2015pattern, liebchen2017collective} are obtained.  The transition from isotropic to polar phase among a 2D populations of self-aligning agents with no excluded volume interactions, is observed in the seminal works, e.g. \cite{vicsek1995novel, toner1995long}. The more complete understanding regarding the transition came much later \cite{gregoire2004onset}. Recently we know that the transition from spatially homogeneous isotropic phase to spatially homogeneous polar phase occurs via a phase coexistence with microphase separation (which gives rise to an effective smectic order) \cite{solon2015phase}. There are several studies where the ordered phases and their properties, such as  giant number fluctuations are explored in experiments \cite{nishiguchi2017long, narayan2007long}.

Often in the experiments it is hard to disentangle the effect of steric and aligning interactions. This motivates one to analyze the collective dynamics and spatiotemporal pattern formation within active systems in presence of both. Various complex dynamical patterns, e.g. polar clusters of varied sizes, band-like or lane-like motile, amorphous structures, flowing crystalline clusters etc. can emerge as a result of the competition between self-propulsion, excluded volume effect and self-alignment interactions \cite{martin2018collective, shi2018self}. It has been claimed in \cite{sese2018velocity} that velocity alignment in such active systems can promote MIPS by reducing the effect of rotational diffusion. Though according to the studies in \cite{weber2014defect, shi2018self}, structural order can also be reduced to a smaller region of the full phase diagram due to aligning interactions. In recent experiments,  in one hand we have evidence showing that aligning interaction can interrupt full phase separation via MIPS \cite{van2019interrupted}, on the other hand it has been shown that a polar flock with excluded volume interaction can undergo a freezing transition in bulk being assisted by MIPS \cite{delphine2019freezing}.

There are remarkable effects, observed within the collective dynamics of self-propelling units, where the confining boundary plays a major role. For example, accumulation of active particles close to the boundary, often leads to non-Boltzmann density profiles which in some simpler cases can be calculated exactly \cite{tailleur2009sedimentation, tailleur2008statistical}. It signifies the out-of-equilibrium character of the system. When an active particle faces a boundary, it cannot turn back from it unless its velocity points away from the boundary. Therefore they are often stuck to it. If there are many such particles, they can come to the boundary one after the other and accumulate. This makes the turning back from the boundary even harder for the particles in front. This effect is observed in varied set-ups, for example --- \cite{bricard2015emergent, vladescu2014filling, wensink2008aggregation, elgeti2009self, volpe2011microswimmers, wysocki2015giant}. The steric interactions between the boundary of the trap and the particles are enough for their accumulation at the boundary without the need for hydrodynamic interaction \cite{elgeti2013wall}. Though the hydrodynamic flows close to the boundary are essential to establish crystalline order within the accumulation resulting from flow induced phase separation (FIPS) \cite{singh2016universal, thutupalli2018flow}. The stability of the order depends on the geometry of the flow fields which essentially depends on the boundary conditions. Inter-particle as well as particle-wall hydrodynamic interaction both can be important for phase separation in a dense, active, confined colloidal systems \cite{zottl2014hydrodynamics}.  Apart from crystallisation, in the presence of confinements, other complex patterns, such as active fluid pumps and vortices are also formed within various confined, simulated active systems (e.g. active Brownian particles and run and tumble particles \cite{nash2010run, hennes2014self, menzel2016dynamical})  and in real active systems (e.g. swimming bacteria, colloidal rollers, vibrated polar discs etc.\cite{wioland2013confinement, lushi2014fluid, bricard2015emergent, kudrolli2008swarming, deseigne2012vibrated}.   

Here we focus on a subtle but simple way to introduce structural order (hexagonal) in addition to the polar order, already established within a motile cluster of confined, self-propelling, soft, circular discs interacting with each other by velocity alignment interaction. No long ranged interaction of hydrodynamic origin is assumed among the particles. It may be noted here that in many cases hydrodynamic interactions do not produce any qualitative difference in the self organisation of active particles (e.g.-\cite{kaiser2017flocking} ).The structural order is established without affecting the motility and polarity of the cluster. The mean packing fraction of the system is kept as low as 15{\%} of the total surface area. The trap to confine the particles and  the inter-particle steric repulsion both are spherically symmetric. In order to develop the structural order within the cluster we will exploit the interplay among their self-propulsion, aligning interactions and boundary effects, in particular, the steepness of the trap boundary.

\section{Main results}

Before going into the details of the model and the results, in this section we will state our main findings briefly. We explore the influence of the confinement on the collective dynamics of active particles to control their organisation. In general, the interplay among activity, self-aligning interaction and steric repulsion offers complex, collective dynamics with pattern forming, non-equilibrium steady states in bulk. Here our focus will be on the role played by the confining boundaries. We demonstrate the influence of the confinement with a model representing soft, active discs with self-aligning interaction and confined within a circular trap. Structural (hexagonal) as well as directional (polar), both order can be established in the system. The phase diagram of the system with respect to both, in the parameter space of activity, boundary properties for a fixed strength of the fluctuations in self-aligning interaction are studied in detail. Through out the study we have kept the packing fraction low ($\simeq$15 {\%} of the surface area) and constant. We find that the fluctuations having strength beyond a threshold value, can destroy both the order. More importantly, we also find that below the threshold, though the polar order is stable with respect to the steepness of the trap, but the structural order is unstable with respect to that. 

\begin{center}
\begin{figure}[h]
\includegraphics[scale=0.50]{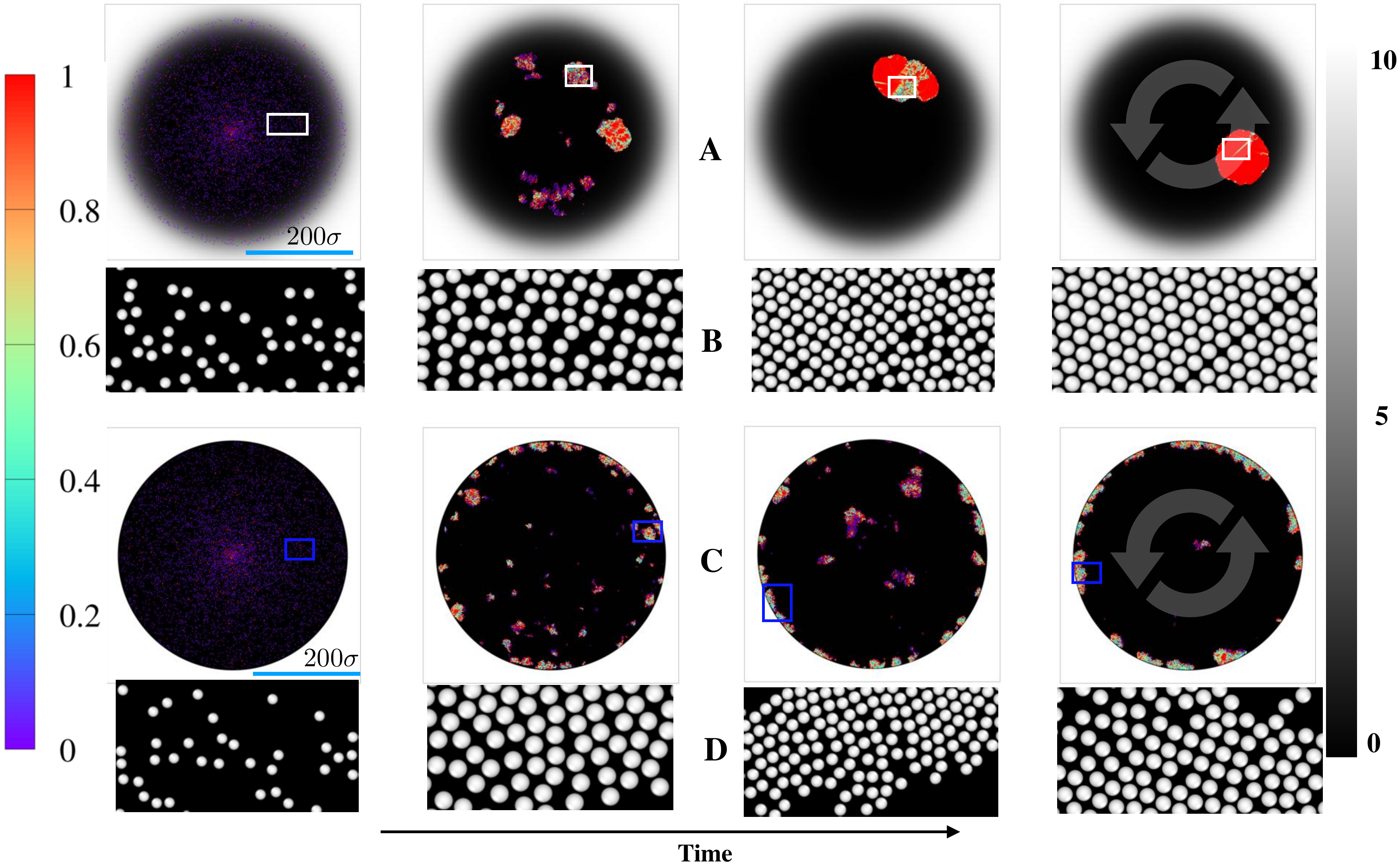}
\caption{{\bf{Structural ordering with low steepness}} ---(A): 4 snaps along time (black arrow) showing configurations with structural ordering measured by the local hexagonal order parameter (defined later) which is unity when all the particles have six neighbours around it forming regular hexagon. The local order parameter for every particle is manifested by the colour of the particles. Higher the value of the local order parameter of a particle, it is in more ordered region of the cluster and vice versa. The trap potential is plotted in grey scale. The potential is non-dimensionalized by the {\it active energy} required by a particle to propel through a distance same as its diameter. Blue scale bar in the left most snap show the dimension of the system. Round grey arrows in the right most snap shows direction of rotation of the cluster. The white boxes are areas of the clusters zoomed in panel B. (B): Zoomed configurations exactly below the corresponding snaps of the full system in (A). The Peclet number for the particles here is $\sim 200$. It implies that the self-propulsion speed of a particle here is $200$ times than the speed related to its diffusion. The dimensionless steepness parameter (defined later) of the trap is $\sim 0.05$.\\ 
{\bf{No structural ordering with high steepness}} --- (C): 4 snaps along time (black arrow) showing configurations without structural ordering, measured by the local hexagonal order parameter. The trap potential is plotted in grey scale and the value of the order parameter for every particle is manifested by the colour of the particles. Blue scale bar in the left most snap show the dimension of the system. Round grey arrows in the rightmost snap shows direction of rotation of the cluster. The blue boxes are areas of the clusters zoomed in panel D. (D): Zoomed configurations exactly below the corresponding snaps of the full system in (C). The Peclet number for the particles here is $\sim 200$ and the dimensionless steepness parameter (defined later) of the trap is $2.5$.}
\label{MainFig}
\end{figure}
\end{center}

\begin{center}
\begin{figure}[h]
\includegraphics[scale=0.50]{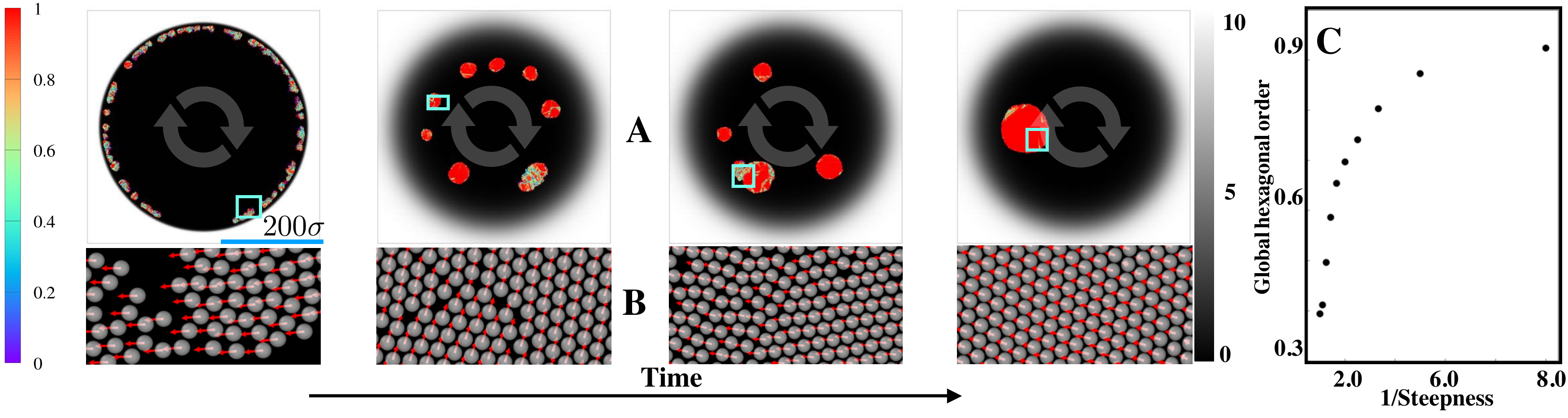}
\caption{{\bf{Controlling self-organisation by steepness}}: (A): 4 snaps along time (black arrow) showing configurations with structural ordering due to quasistatically varying steepness of the trap. The trap potential is plotted in grey scale. The potential is non-dimensionalized by the {\it active energy} required by a particle to propel through a distance same as its diameter. The value of the local hexagonal order parameter (defined later) for every particle is manifested by the colours of the particles. Higher the value of the local order parameter of a particle, it is in more ordered region of the cluster. Round grey arrows are direction of rotation of the cluster. The cyan boxes are areas of the clusters zoomed in panel B. (B): Zoomed configurations exactly below the corresponding snaps of the full system in (A). Red arrows imples velocities of the particles. (C): The structural order within a cluster quantified by global hexagonal order parameter (defined later) , increases with inverse of the steepness of the trap. The Peclet number for the particles here is fixed at $\sim 200$.}
\label{MainFig1}
\end{figure}
\end{center}

We demonstrate this by simulating the system from same random initial configuration but with two different steepness of the circular trap (everything else e.g., depth of the trapping potential, packing fraction, activity, self-aligning and steric interactions etc., are kept same for the two), which finally produces two differently organised steady states --- for low steepness the final state is structurally ordered and for high steepness there is no structural order (Fig[\ref{MainFig}]). In both the cases particles accumulate close to the boundary and form dynamic, polar clusters that rotate along the boundary of the circular trap. The size of the clusters of the active particles depends on the length scales set by the trap. To emphasise  the influence of the steepness of the trap boundary further and to achieve a control on the degree of the structural order, one can also start from the polar but structurally disordered phase and reduce the steepness slowly (quasistatically) over time to develop the order in addition to the polar order within the cluster. This can be used as a strategy to control the degree of the order within the polar cluster of self-propelling and self-aligning active particles in a confinement.  In Fig[\ref{MainFig1}] we have shown that how the structural order varies with the steepness of the trap.   

After identifying the relevant parameter spaces for various phases in the phase diagrams we explore the kinetics of the ordering. We find that the transition from the isotropic to the polar phase in our system resembles the same within a population of Vicsek agents. The transition is robust even in the presence of the confinement and steric interactions. With decreasing steepness of the trap, we find that the structural order can emerge within the self-organised polar phase of the active particles, keeping its polar order intact. The structural order is the outcome of the interplay among the spatially non-uniform repulsion from the boundary, self-propulsion and self-aligning interactions present in the system. When the boundary of the trap is very steep, particles within the clusters are rotating along the boundary.  The interface between the cluster and the trap is smooth. At any point of time, the velocity direction of individual interfacial particle points almost along the tangent of the circular trap at that point. With reducing steepness, while rotating, interfacial particles start to bounce at the boundary of the trap repeatedly. Consequently the smooth, circular interface turns into rough by creating ripples. With further decrease of the steepness, the ripples grow and the polar cluster is fragmented into smaller ones.  Smooth fronts or interfaces being unstable, can break into clusters due to different types of dynamical instabilities. For example, in a suspension of colloidal micro-rollers fingering instability, originated from purely long-ranged hydrodynamic interactions, can break smooth shock front into small clusters \cite{driscoll2017unstable}. In the system of our concern, after fragmentation, the smaller clusters also bounce repeatedly at the trap boundary with low steepness. Consequently a force couple acts within the clusters to make individual cluster rolling while rotating along the boundary. Due to rolling, particles within a cluster are facing the confining force from all directions. Eventually the clusters become compact enough to support the structural order, keeping its polarity and collective motion intact. These polar and structurally ordered clusters can merge together to become a single cluster or can collide with each other. Though these events can drastically destroy the orders within a cluster, but it is a transient phenomena. After one such event, if a cluster is allowed to relax to a steady state without facing further collision or merging events, eventually it regains both structural and polar order. It implies that the orders, which are developed within the clusters, are robust against such events. Hence we obtain a simple protocol to control the degree of structural order within a polar cluster which can be implemented in experiments. Pertinent issues regarding crystalline order within confined active particles are raised and investigated in \cite{briand2018spontaneously}. There, initially self-propelling and spontaneously aligning hard discs are allowed to form a close-packed configuration with perfect hexagonal order within a hexagonal arena. Removing particles from the configurations causes condensation of shear along the stacking faults, that leads to large scale shear flow within the system. Surprisingly both structural and polar order can survive against the flow, providing us flowing, polar crystal. But the packing fraction used in the study is very high (~$\simeq$81{\%}) compared to the study here. Consequently, we will see that the kinetics of ordering here is fundamentally different from \cite{briand2018spontaneously}.  We have also explored the stability of the structural order when the confinement is suddenly turned off and showed that the decay of the order can follow power law over time. The power can vary over time intervals, indicating multiple regimes of the decay, crossing over one by the other. The details of the results will be provided later.

%In particular, together they generate a torque within the polar cluster of active particles, which eventually compactify the cluster that gives rise to the order in the presence of the steric repulsion among the particles. The details of the results will be provided later. 

%In the figures [] we have shown that with low steepness trap boundary how the structural order emerges from position-wise random configurations (polar as well as non-polar). If the steepness is high, we have also shown that no such structural order develops.       

\onecolumngrid

\twocolumngrid

\section{Model}
\label{sec_model}

%\subsection {The Active Particles} 

We consider a collection of $N$ self-propelling particles at $\{{\bf{r}}_i\}$ with velocities $\{{\bf{v}}_i\}$ where $i=1,2,..., N-1, N$. They are confined within a two dimensional circular trap modelled as a continuous, spherically symmetric function of $r_i=|{\bf{r}}_i|$ as,

\begin{eqnarray}
U(r_i)=U_0(\tanh(q(r_0-r_i))+1)
\end{eqnarray} 
such that $2U_0 \ge U(r_i)\ge 0$. We consider that the boundary of the trap is at $r_0$, where the potential felt by a particle is : $U(r_0)=U_0$. The energy of the individual self-propelling particle  remains always lower than $U_0$ such that they can never cross the circular barrier. The parameter $q$ controls the steepness of the boundary. The extent of the repulsion from the boundary towards the centre of the trap can be quantified by the length scale $l=q^{-1}$. It is small if $q$ is large and vice versa. Though the magnitude of the repulsive force from the boundary will increase as $q$ becomes larger. Therefore, large $q$ limit is close to the hard wall limit where the layer of the particles closest to the boundary faces a very high repulsive force but the layers afterwards face a very small repulsion directly from the wall. On the other hand, when $q$ is small, the potential gradient and consequently the force gradient due to the boundary become smoother, making it soft.  In case of the soft boundary, the extent of the repulsive force from the boundary is larger. In this case, not only the layer of the particles closest to the boundary but the layers behind can also feel the repulsion from it. Though its magnitude decreases gradually to zero as one goes towards the centre of the trap from its boundary. In this work, we focus on how the gradient which is controlled by a single parameter $q$ of the potential, affects the self-organisation of self-propelling particles confined within it. Through out the work we consider $qr_0>>1$ such that the influence of the boundary on the particles is considerable only when they are close to it and become negligible as we go far from it i.e. in the central region of the trap.

While propelling with constant force $f_0$, an individual particle {\it{try}} to follow its neighbours within a certain circular area  of radius $r_{int} << r_0$ surrounding the particle. In order to do that the particle follows Vicsek-like aligning interaction\cite{vicsek1995novel, weber2014defect}. The particle decides its direction of active force ${\bf{\hat{n}}}_i$ to be aligned with the average velocity direction of its neighbours including itself. This local alignment is not perfect due to the presence of an angular noise $\theta$ randomly chosen between $0$ and ${2\xi}{\pi}$ at every time step. Here $\xi$ is a dimensionless parameter that lies between $0$ and $1$ and defines the strength of the angular noise. One can express the alignment interaction as,

\begin{eqnarray}
\hat {\bf n}_i=R_{\xi}(\theta)\circ\left(\frac{1}{N_{S_i}}{\sum_{k\in S_i}
    \hat{\bf n}_k}\right)
    \label{Vicsek}
\end{eqnarray}
where $\hat{\bf{n}}_k={\bf{v}}_k/{|{\bf{v}}_k|}$, $S_i$ is the neighbouring circular area of $i$th self-propelling particle, within which there are $N_{S_i}$ particles that include the neighbours of $i$th particle and $i$th particle itself. They are indexed with $k=1,2,....,N_{S_i}$. $R_{\xi}(\theta)$ is the operator that rotates (denoted by $\circ$) the average direction of the neighbours within $S_i$ by the randomly chosen angle $\theta$ to model the imperfection in the alignment. Instead above, one may also consider continuum version of Vicsek dynamics \cite{farrell2012pattern, liebchen2017collective, martin2018collective} though the qualitative features of the system will remain same \cite{chepizhko2010relation}.

Within a distance smaller than $r_{int}$, the particles are also repelling each other by a spherically symmetric force derived from a soft, continuous potential $V_{wca}$: Weeks-Chandler-Andersen (WCA) interaction \cite{weeks1971jd} to account for the effect of excluded volume interaction between the particles,
\begin{eqnarray}
V_{wca}&=&4\epsilon\left(\frac{\sigma^{12}}{r_{ij}^{12}}-\frac{\sigma^{6}}{r_{ij}^6}\right)+\epsilon \phantom{xxx}
{\text{for}}\phantom{x} r_{ij}<2^{1/6}\sigma \nonumber \\
&=& 0. \phantom{xxxxxxxxxxxxxxxx}{\text{elsewhere}}
\end{eqnarray}
Here $r_{ij}$ is the distance between i-th and j-th particles and $\epsilon$ determines the strength of the repulsive potential between them. The length scale involved in the repulsive potential is $\sigma$ which is the measure of the size of an individual particle. We have considered $\sigma$ as the unit of length. We consider $r_{int}=2\sigma$ to ensure alignment even when particles are very close to each other.  Here we assume that the alignment is of non-steric origin which may be realised by the systems having spherically symmetric vibrated disks \cite{scholz2017velocity} or active colloids under magnetic or long-ranged hydrodynamic interactions respectively. It can also be mentioned here that the model described here captures the essential features of active systems such as migratory cells \cite{stichel2017individual}, where both the effects of excluded volume and alignment interactions are relevant. 

At time $t$, if the position and velocity of $i$th particle are denoted by ${\bf{r}}_i(t)$ and ${\bf{v}}_i(t)$, its dynamics can be written within Langevin paradigm as, 
\begin{eqnarray}
m\frac{d{\bf{v}}_i}{dt}=-\gamma{\bf{v}}_i-{\bf F}_i +f_0\hat
{\bf{n}}_i+\sqrt{2\gamma K_BT}{\bf{\eta}}_i
\label{EqnOfMotion}
\end{eqnarray}
where $m$ is the mass of a particle, $\gamma$ is the friction coefficient due to the surrounding fluid and
${\bf{\eta}}_i$ is the thermal noise, i.e. $\langle\eta_i\rangle=0, \langle\eta_i(t_1)\eta_j(t_2)\rangle=\delta_{ij}\delta(t_1-t_2)$.  We consider mass of the particle to be unity and $\gamma$ to be fixed at high value ($\sim 10$) such that for unit $\epsilon$ and $\sigma$, $\sqrt\epsilon/\gamma\sigma << 1$ which is motivated by the systems with high viscosity (e.g. colloidal suspensions, bacterial suspensions etc.). The force ${\bf{F}}_i$ is the superposition of all the forces on $i$th particle, provided by the trapping potential $U$ and collisions with other particles via WCA interaction, i.e. ${\bf{F}}_i=-\nabla_iU+\sum_j {\bf{F}}_{ij}^{wca}$. $f_0$ is the magnitude of self-propelling force, the direction of which is decided by Eq.[\ref{Vicsek}]. The strength of the thermal noise is related to the friction coefficient and temperature of the surrounding fluid by fluctuation-dissipation relation which ensures the equilibrium of the system in absence of self-propulsion. We consider the system to be far from equilibrium due to self propulsion where the effect of passive, thermal diffusion is small. Therefore through out the calculation we consider $K_BT/f_0\sigma \leq 10 ^{-4}$. 

In similar systems, role of the defects on the verge of the transition from fluid like disordered phase to ordered phases (for example, hexagonal and polar phases) have been explored in bulk \cite{weber2014defect}. Recently, while varying different parameters, it has been shown that similar model can also accommodate other complex steady states in bulk, e.g. finite-sized polar clusters, band-like or lane-like highly dynamic structures and their hybrids \cite{martin2018collective}. Collective dynamics of self-propelling chiral entities are investigated by similar models with Vicsek-like alignment combined with active torque \cite{liebchen2017collective}. Here we will explore the collective dynamics of such self-propelling, mutually aligning, finite-sized particles under the influence of a circular confinement. We will show that the steepness of the trap has major impact on the self organisation of such active particles. In particular, for a fixed packing fraction of confined active particles, depending on the Peclet number and the strength of the rotational noise, by tuning steepness of the trap one can induce structural disorder-order transition into them.  We will keep the packing fraction $\phi={N\sigma^2}/{r_0^2}= 0.15$ fixed through out the work to investigate the effects of alignment, excluded volume and confinement in detail.

%At this packing fraction no such transition is observed in the following limiting cases: i) un-confined (but with periodic boundary) where $U_0\rightarrow 0$, ii) passive case with $f_0\rightarrow 0$ and iii) active but no-alignment case with $r_{int}\rightarrow 0$. In any one of these limiting cases the system remains in a non-polar, gas-like disordered state. It ensures that the transitions we observe here are outcome of following three ingredients only: activity, alignment and confinement. 

At this stage it is convenient to identify few non-dimensional parameters that control the collective behaviour of the confined particles. Individual particle is experiencing forces from other particles due to collisions which is of the order of ${\epsilon}/{\sigma}$; it is experiencing thermal force of the order of ${K_BT}/{\sigma}$ due to the surrounding fluid at temperature $T$; then from the boundary of the trap it is facing repulsive inward force of the order of $qU_0$. Finally due to active degrees of freedom, individual particle exerts force $f_0$ to their surrounding. To ensure confinement, here we consider that ${qU_0}/{f_0}, {qU_0\sigma}/{\epsilon}$, ${qU_0\sigma}/{K_BT})\geq 1$. Again, we consider ${\epsilon}/{K_BT}, {f_0\sigma}/{K_BT}$ to be larger than unity such that we can focus on the interplay between steric interaction and activity only. 

For the system here, we define Peclet number as, 

\begin{eqnarray}
\text{Pe}=\frac{f_0}{\gamma\sqrt{K_BT}}.
\end{eqnarray} 
which quantifies the self-propulsion with respect to thermal diffusion by taking the ratio of friction limited active velocity and thermal velocity of the particles. With $\text{Pe}\rightarrow 0$ the system becomes passive (equilibrated with temperature $T$), losing their active degrees of freedom ($\{{\bf {\hat n}}^t_i\}$) and related dynamics. 

In the system of our interest, there are competing aligning and de-aligning interactions. The competition can be quantified by their respective ranges. The ranges of Vicsek-like aligning and simultaneously de-aligning interaction are given by $s_1=\pi r_{int}^2$ and $s_2=\xi\pi r_{int}^2$ respectively. The competition between them can be quantified by the parameter $s_2/s_1=\xi$ that varies between $0$ and $1$. At $\xi\rightarrow 1$ limit the particles can be maximally de-aligned due to rotational fluctuations in the alignment interaction and for $\xi\rightarrow 0$ the opposite happens. One can note here that for alignment interaction to work, particles should {\it{see}} each other, i.e. the average distance between them should be at least $r_{int}$ or less. As we keep the packing fraction low ($\phi=0.15$), they can see each other only after a certain critical value of Pe. We will be back to this point while discussing the phase diagrams in later section. The limit where $\xi\rightarrow 1$, the self-propelling particles resemble active Brownian particles (ABP) \cite{fily2012athermal, cates2015motility, redner2013structure, stenhammar2013continuum, stenhammar2014j, buttinoni2013dynamical} confined within the trap. In 2D, ABP encounter MIPS in bulk at much higher density as compared to here. In the regime of density and Pe we are concerned here, as ABP lack alignment interactions, it will not produce clusters beyond the critical size required for MIPS.

%Next, we identify rotational Peclet number as,

%\begin{eqnarray}
%\text{Pe}_r=\frac{1}{\xi}\left(\frac{r_{int}}{\sigma}\right)^2
%\end{eqnarray} 
%which quantifies {\it{aligning}} interaction with respect to {\it{de-aligning}} interactions i.e. the inter-particle repulsion via WCA and the randomising angular noise $\xi$. With $\text{Pe}_r\rightarrow 0$, the system becomes a collection of active Brownian particles [] trapped within a circular cavity. With $\sigma\rightarrow 0$ (i.e. $\text{Pe}_r\rightarrow \infty$), the particles become Vicsek agents [] confined in the cavity.   

The particles here can behave as effectively soft as well as hard, depending on certain condition. The inter-particle steric interaction between any two particles here becomes increasingly repulsive as they come close and it diverges when their centres overlap i.e. at $r_{ij}=0$. Therefore, the particles can be soft only up to a certain extent and beyond that it becomes extremely hard such that the particles cannot come closer to each other. The hardness of the particles can be parametrised as 

\begin{eqnarray}
\Lambda=\frac{f_0}{\gamma\sqrt{\epsilon}}
\end{eqnarray} 
If $\Lambda > 1$ the particles are soft and if $\Lambda \leq 1$, it is hard.  Here, varying $f_0$ while keeping the other parameters fixed, we have explored hard as well as soft particles.  For very large $\Lambda$ and Pe (in the limit of $\epsilon\rightarrow0, K_BT\rightarrow 0$), the particles are close to Vicsek agents.  In bulk, the homogeneous state of Vicsek agents can be transformed to a polar state when $\xi$ is lower than a critical strength $\xi^c$ via a first-order phase transition and when trapped, they can exhibit different complex dynamical patterns (e.g. swarms) \cite{vasarhelyi2018optimized}.       

Finally, the dimensionless steepness parameter of the circular wall of the cavity can be written as 

\begin{eqnarray}
S=q\sigma.
\end{eqnarray} 
We will show that one can tune $S$ by varying $q$ to control the self-organisation of the particles externally.\\ 
In the next section onwards, after mentioning simulation details, we will present the phase behaviour of the system, with respect to activity (Pe) and the steepness parameter ($S$) of the trap for fixed value of $\xi$ that supports polar order. Next we will discuss the kinetics of phase transformations together with the stability of the ordered and polar phases and finally we will conclude by summarising the work together with its possible experimental realisation.      

%***********************************************

%\section{Simulation Details}

We simulate the system by integrating the equations of motion given in Eq[{\ref{Vicsek}}] and Eq[{\ref{EqnOfMotion}}] simultaneously with open boundary condition. To update velocity and position we follow velocity Verlet algorithm with $dt=10^{-3}$. All the data presented here are tested with total number of particles $N=5 \times 10^3$ and $N=10\times 10^3$ particles to estimate the finite size effect which has turned out not to cause any qualitative change to the data presented here. To analyse the data we have used OVITO \cite{stukowski2009visualization}.     

%***********************************************

\section{Results}

\subsection{Order Parameters}
We begin by analysing the steady states of the system obtained by varying Pe, and $S$ at fixed packing fraction of the particles. The steady states of the system are characterised by their spatial and directional organisations. The spatial organisation of the particles is quantified by the local (at the level of individual particle) hexagonal order parameter, defined as   

\begin{eqnarray}
\psi_{6,i}=\frac{1}{N_i}\sum_j\exp({\bf{i}}6\theta_{ij})
\end{eqnarray}  
where ${\bf {i}}=\sqrt{-1}$ and $N_i$ is the number of topologically (Voronoi) nearest neighbours of $i$th particle. The summation is over all such neighbours of $i$th particle and $\theta_{ij}$ is the bond angle between $i$th and $j$th particles with respect to some arbitrarily chosen reference frame. For a perfect triangular lattice in 2D, $Q_6^i=|\psi_{6,i}|$ is unity. With disorder its value reduces from unity and for a maximally disordered configuration it is zero. For a given configuration of particles, with $Q_6^i$ one can quantify local order and disorder by identifying ordered zones and defects in between.   
To quantify the global positional order of the system in a steady state we consider global hexatic order parameter as $Q_6=\langle |\langle\psi_{6,i}\rangle_i|\rangle_{\text{(r,t)}}$ where $\langle .\rangle_i$ denotes the averaging over all particles and $\langle.\rangle_{(r,t)}$ denotes averaging over realisations and time in steady state. The state of the system having $Q_6$ above (below) a certain threshold is ordered (disordered).
 
 The local polar order of individual particle can be quantified by the angle made by $\hat{\bf{n}}_i$ with respect to an arbitrarily chosen reference frame and to quantify the global polar order, following \cite{vicsek1995novel, weber2014defect, martin2018collective} we define the polar order parameter as,

\begin{eqnarray}
{\mathcal{P}}=\left | \frac{1}{N}\sum_{i}\hat{\bf{n}}_i\right |
\end{eqnarray} 
where ${\mathcal{P}}\equiv|{\bf {P}}|=|P_r{\hat {r}}+P_{\theta}{\hat {\theta}}|=\sqrt{P_r^2+P_{\theta}^2}$ in plane polar coordinate. When all the particles are propelling being perfectly aligned with each other, $\mathcal{P}=1$ and when they are moving randomly, $\mathcal{P}=0$.  In a steady state, if the time and realisation averaged value of $\mathcal{P}$ goes beyond a certain threshold, the state is polar and otherwise non-polar. 

The polar state here is the reminiscent of the flocking state of Vicsek agents \cite{vicsek1995novel}, which can be stable when the value of $\xi$ is below a threshold value. In this state, when the particles encounter the boundary of the trap, the radially outward component of the force exerted by the particles to the trap is balanced by radially inward repulsion from the boundary. Also in the polar state, as $\xi$ is low, it is more probable for a particle to remain close to the boundary than to turn around from it \cite{tailleur2009sedimentation}. Thus the particles in the polar state accumulate close to the boundary and flock together along it, giving rise to self-rotating clusters in the trap. The sense of rotation is spontaneously chosen. In a similar set up where hydrodynamic and electrostatic interactions are also taken into account, formation of other complex structures such as vortices are also reported \cite{bricard2015emergent}. They are not finite size effect but genuine, macroscopic, non-equilibrium phases of active matter in confinement \cite{caussin2015collective}. The major concern of our work is the emergence of structural (e.g. hexagonal) and polar order within such dynamic clusters.  

Due to circular symmetry of the boundary, in a highly polar steady state, we find that $P_r << P_{\theta}\simeq {\mathcal{P}}\rightarrow 1$. In the state where the particles are moving randomly, both $P_r$ and $P_{\theta}$ are very small, close to zero, causing ${\mathcal{P}}\rightarrow 0$. Therefore in the system of our concern, emergence of polarity can equivalently be quantified by $P_{\theta}$ alone as $P_r$ is negligibly small in both polar and non-polar states.

%where $N_S$ is the number of neighbours within the circular area $S$ around $i$-th particle, as before. If at certain time, all neighbours are randomly oriented or there is no neighbour of $i$th particle, then for that particular particle, at that time ${\mathcal{P}}_i=0$. If the neighbouring particles are all perfectly aligned with $i$ th particle, then ${\mathcal{P}}_i=1$. To convert from local to global order parameter, above order parameters should be averaged over all the particles,

%\begin{eqnarray}   

\subsection {Phase Behaviour}

\begin{figure}[ht]
\centering
\includegraphics[width=0.80\columnwidth]{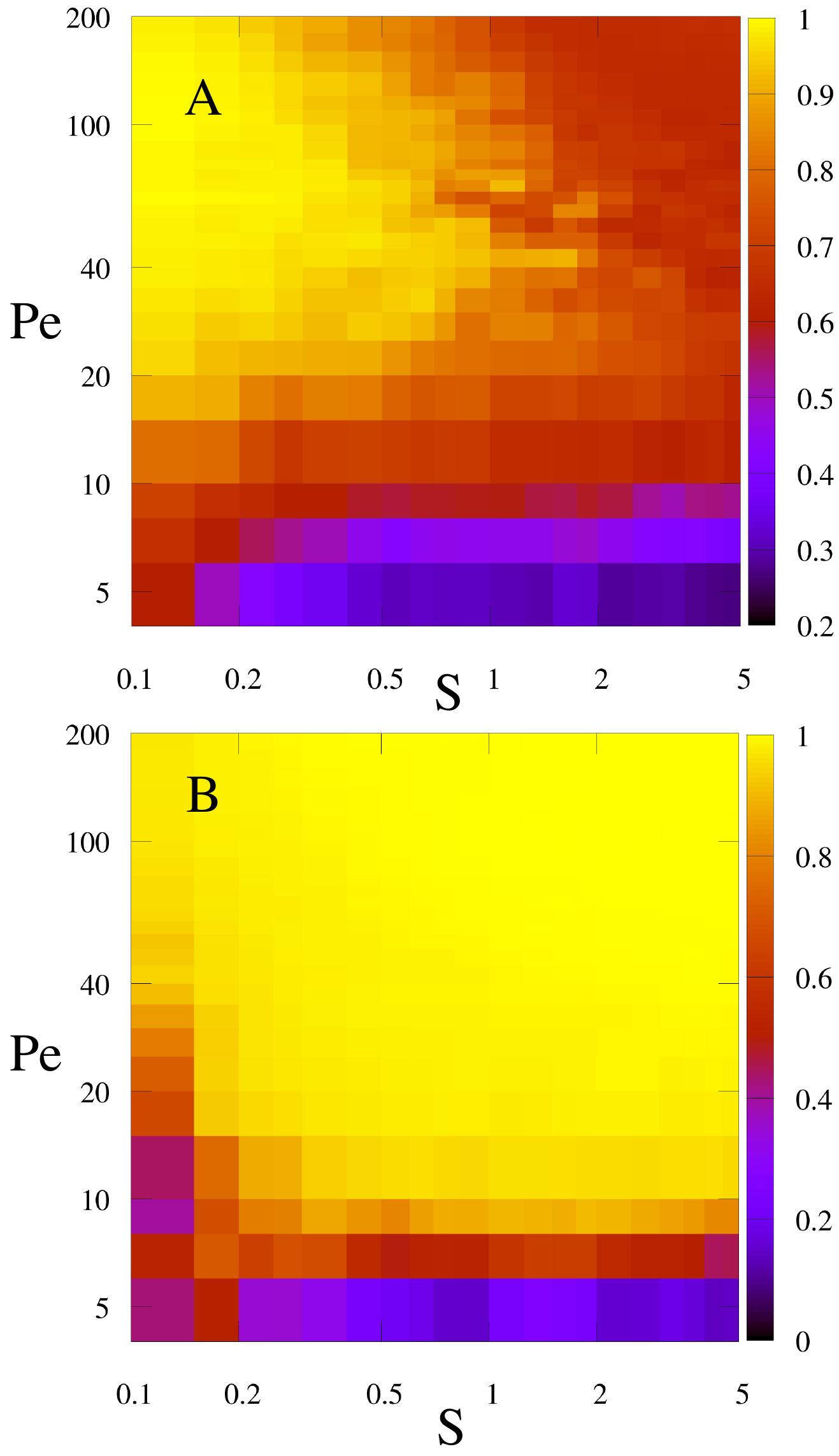}
\caption{{\bf{Phasediagram}} : (A) normalised hexagonal order ($Q_6$) is plotted in log-scale along Pe-S plane for $\xi=1/2\pi$. When $Q_6\leq 0.3$, particles cannot form clusters having length scale beyond the size of the particles. On the other hand, well developed hexagonal order exists within the configurations having $Q_6\geq 0.80$. In between, particles form clusters but the order is not well developed within them. (B)  polar order $\mathcal{P}$ plotted in log-scale along Pe-S plane with $\xi=1/2\pi$. The polar order is not well developed within the configurations having $\mathcal{P} \leq 0.5$. Elsewhere it is well developed.}
\label{Phase1}
\vspace{-3mm}
\end{figure}

We analyse the phase behaviour of the system by calculating $Q_6$ and $\mathcal{P}$ in steady states with different values of Pe and $S$. We fix $\xi=1/2\pi$. In steady states we calculate $Q_6$ to estimate the structural order that varies along $\xi=1/2\pi$ plane of a 3D phase diagram having three axis: Pe, $ S \phantom{x} \text{and} \phantom{x} \xi $. This variation is plotted in Fig.[\ref {Phase1}A]. Similarly in steady states we also calculate $\mathcal{P}$ to estimate the polar order that varies along $\xi=1/2\pi$ plane Fig.[\ref {Phase1} B]. Below we will discuss them one after the other.

{\bf{(a)}} According to Fig.[\ref {Phase1}A], we will discuss various phases with varying $S$ but fixing different ranges of Pe. For all $S$ but at low Pe (range : 1 $\le$ Pe $\le$ 8 ), we find that the particles are randomly distributed through out the trap. In this regime, thermal fluctuations dominate over activity, prohibiting accumulation close to the boundary. The value of $Q_6$ is quite low ($Q_6 \le 0.4$). 

%Typical configuration of the particles in this parameter space is given in Fig[].  

As we raise Pe (range : 8 $\le$Pe$\le$ 15), self-propulsion starts to dominate over thermal fluctuations and the particles accumulate close to the boundary forming clusters. The corresponding value of $Q_6$ becomes higher than the earlier. It remains between 0.4 and 0.55. The degree of ordering is still low. There is a remarkable difference between the shape of the clusters at $S$ lower than a threshold value $S_c$ ($S_c \simeq 0.70 $) and clusters at $S > S_c$. At higher steepness the clusters wet the circular wall of the trap i.e. the contact angle between the layer of the particles and the boundary of the trap is very small \cite{de2013capillarity} . On the other hand, if the steepness is reduced, the clusters become somewhat round, increasing the contact angle between the clusters and the wall of the trap. 

%As we raise Pe (range : ), activity starts to dominate over thermal fluctuations and the particles accumulate close to the boundary forming clusters. The corresponding value of $Q_6$ becomes higher than the earlier (It remains between 0.35 and 0.50). It is apparent from the typical configurations given in Fig[] that there is a remarkable difference between the clusters at lower ($S \leq 1.0 $) and higher steepness ($S \geq 1.0 $). At higher steepness the clusters wet the circular wall of the trap i.e. the contact angle between the layer of the particles and the wall of the trap is very small. On the other hand, we find that lowering the steepness promotes de-wetting [] where the clusters becomes somewhat round, increasing the contact angle between the clusters and the wall of the trap.  

Next, when Pe is increased further (range : 15 $\le$ Pe $\le$ 40), we find a clear order-disorder transition, together with the shape transformation of the clusters as mentioned before, when we vary $S$ beyond a threshold value $S_c$. In this range of Pe, the steady states for $S < S_c$ ( $S_c \simeq 1.0$), the clusters are round in shape and they have polycrystalline order with dynamic grain boundaries \cite{godreche1991solids}. The average value of $Q_6$ in this regime (around $S \rightarrow S_c^{-}$) can become as high as 0.85. While increasing $S$, beyond $S_c$, the clusters become strip-like, wetting the boundary of the trap. They loose their structural order. Consequently, $Q_6$ decreases up to 0.55. 

%Next, when Pe is increased further (range : 20 $\le$Pe$\le$ 40), we find a clear order-disorder transition as we vary $S$ over a broad range. In this range of Pe, the steady states for low $S$ ( $\leq 1.0$), the clusters are round in shape and they are polycrystalline with dynamic grain boundaries [] sandwiched between differentially oriented single, triangular crystalline zones (see typical configuration in Fig[]). The average value of $Q_6$ in this regime is as high as 0.70. While increasing $S$, beyond $S=1.0$, the clusters again become strip-like, wetting the wall of the trap. They loose their positional order. Consequently, $Q_6$ decreases up to 0.50. Typical configuration for this regime of the phase diagram is given in Fig[]. 

Next, Pe is increased even further (range : 40 $\le$Pe$\le$ 200). In this regime there is always an order-disorder transition if one goes beyond $S=S_c$ while increasing $S$. Interestingly we find that in this range, as Pe becomes gradually higher, $S_c$ becomes gradually lower. At Pe$=$200, the transition happens around $S_c \simeq 0.5$ whereas at Pe$=40$ it was taking place around $S_c=1.0$. From lowest value of S considered here is $0.1$. In this range of Pe, from the lowest value of $S$ up to $S = S_c$ we obtain round-shaped cluster with hexagonal order as earlier and beyond we obtain strip like clusters along the circular wall with reduced structural order ($Q_6 \simeq 0.55$). This indicates that for very low and very high values of Pe, degree of the structural order within the clusters is low. Only the values of Pe in between (range : 20 $\le$Pe$\le$ 200) can facilitate the ordering if the steepness of the trap is below $S_c$. \\ 
%Next, Pe is increased even further (range : 40 $\le$Pe$\le$ 200). Interestingly we find that in this range, as Pe becomes higher gradually, the order-disorder transition happens at lower $S$ ($\simeq 0.70$) than earlier. Up to $S \simeq 0.70$ we obtain round-shaped cluster with polycrystalline configuration (see Fig []) as earlier and beyond we obtain strip like clusters along the circular wall with reduced positional order ($Q_6=0.50$). This implies that for very low and very high values of Pe, hexatic order within the clusters is low. Only the moderate values of Pe (range : ) can facilitate the ordering if the steepness of the trap is low enough. 

{\bf{(b)}} From the phase behaviour with $\mathcal{P}$ (Fig.\ref{Phase1}B) we find that two distinct regions separated by a boundary. For Pe$ > 10$ and $S > 0.2$ we find the system is polar ($\mathcal{P} \simeq 0.9$). To reach to the polar phase, the particles have to $\it{see}$ each other. As the packing fraction is low, the Pe of the particles required to interact with each other should be beyond a threshold which is around $10$ here.  The polar clusters rotate along the circular boundary of the trap with spontaneously chosen direction (clockwise/anticlockwise). On the other hand for Pe$ < 10$ the particles are moving randomly and therefore we obtain isotropic phase with $\mathcal{P} \le 0.2$.  In this regime thermal fluctuation plays a dominant role which prevents clustering. One can note here that the transition from isotropic to polar phase is almost independent of the steepness of the trap, unless $S < 0.2$. For very low steepness the transition happens at higher Pe. \\

\subsection {Kinetics of Ordering}

From now onwards we will focus on how the structural order emerges in the system with Pe $\ge 20$ and when $S$ is decreased from the values larger than $S_c$ up to the values less than $S_c$, keeping $\xi$ fixed at $1/2\pi$. The value of the packing fraction $\phi$ is also fixed at $0.15$. At this parameter space the system becomes polar. From the phase diagram before it is apparent that even if the steepness of the trap is varied over a broad range, it is not able destabilise the polar order rather while lowering, it introduces another order---the structural (hexagonal) order---in addition to the polar order. The onset of such order is analysed below.    

\onecolumngrid

\begin{center}
\begin{figure}[h]
\includegraphics[scale=0.50]{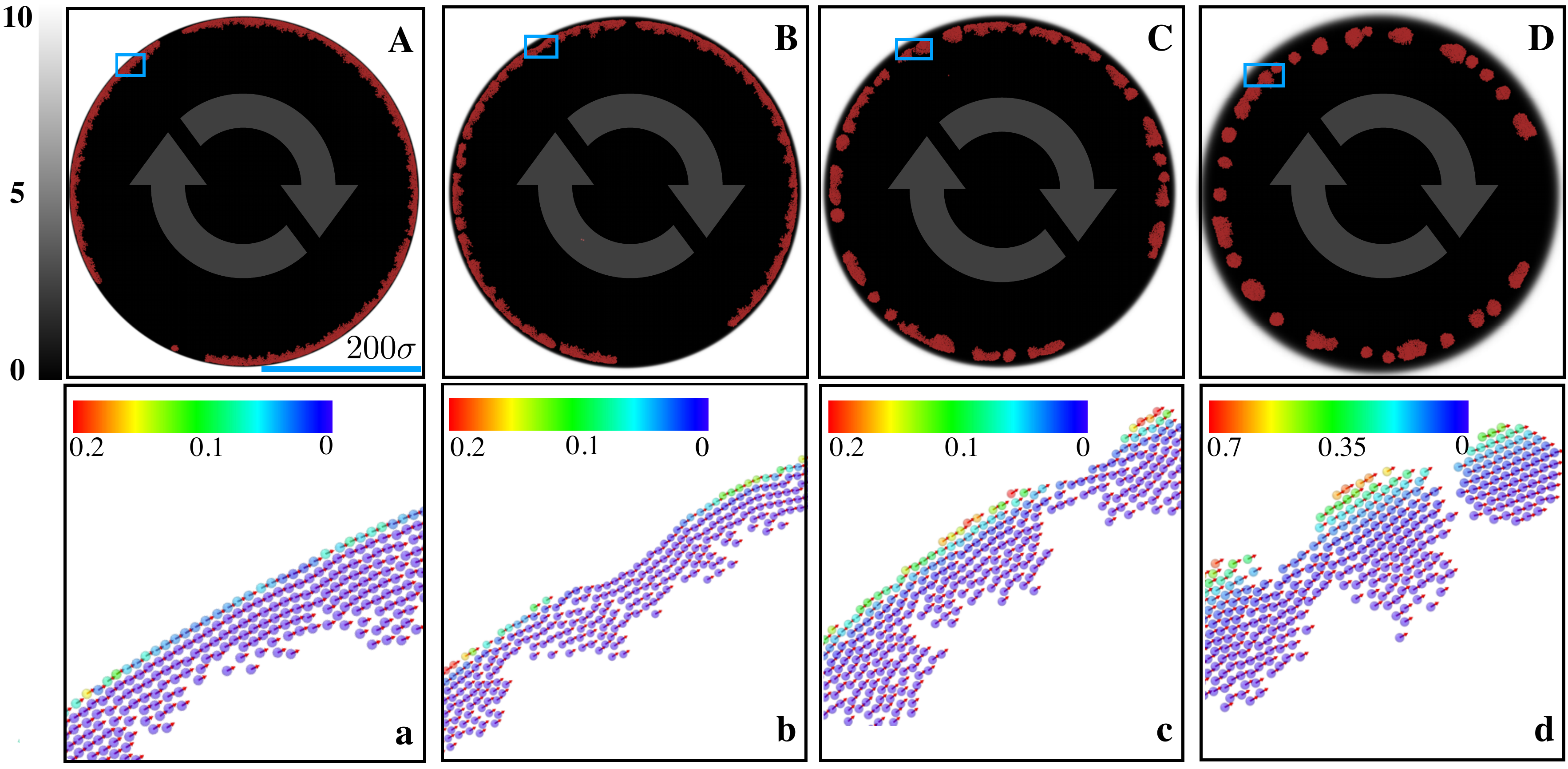}
\caption{{\bf{Fragmentation}}: (A) Self-propelling particles (in brown) accumulated at the circular boundary of the trap with $S=2.5$, forming strip-like polar cluster. Grey arrow : direction of rotation of the clusters. Grey scale : potential strength. The potential is non-dimensionalized by $f_0\sigma$.The configuration within the blue box is zoomed in (a). (a) Positions of the particles are plotted with their directions of velocities (red arrows). Colour scale : strength of the trapping potential faced by individual particle. The interface between the particles and the trap boundary is smooth. 
Next, in (B), (C) and (D), positions of the self-propelling particles and the trapping potential are plotted in the same way, as in (A). Similarly, in (b), (c) and in (d), the zoomed positions of the particles, together with their velocity arrows and the trapping potential faced by the individual particles are represented in the same way as in (a).
(B) All the parameters are same as in (A) except, the steepness of the trap $S=0.95$. (b) As the steepness decreases ripples appear at the interface. (C) Here $S=0.53$. All other parameters are same as earlier. (c) Ripples grow as the steepness decreases even further. (D) Here $S=0.26$. All other parameters are same as earlier. (d) Fragmentation occurs as the steepness is decreased further. }
\label{MainFig2}
\end{figure}
\end{center}         

\begin{center}
\begin{figure}[h]
\includegraphics[scale=0.50]{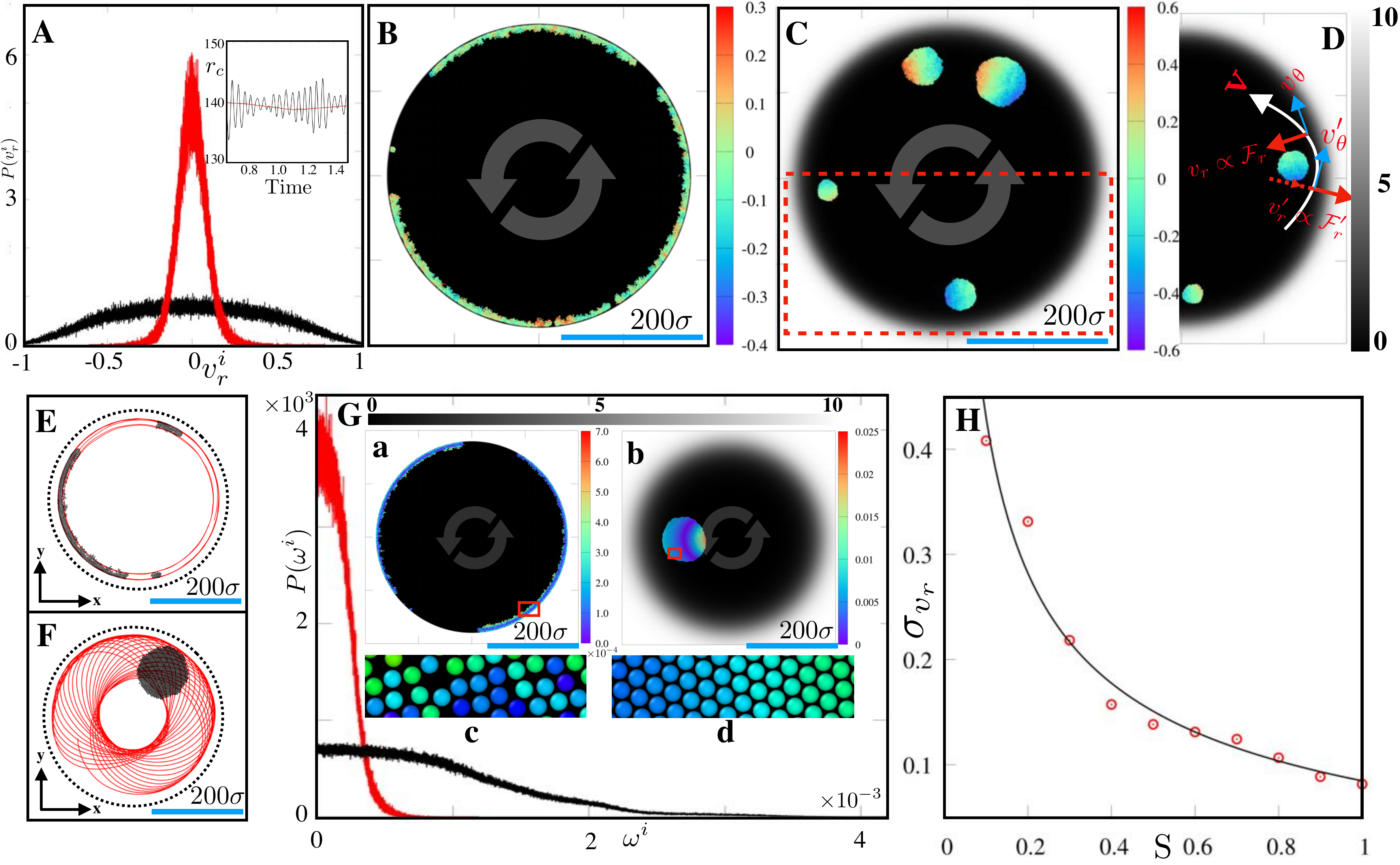}
\caption{{\bf{Rolling and structural ordering }}: (A) Probability distribution of radial component of velocities of the particles. Red : non-bouncing cluster. Black : bouncing cluster. (A, Inset) Black : position vs. time for centre of mass of bouncing cluster. Red : position vs. time for centre of mass of non-bouncing cluster. (B) Cluster of particles spreads along the boundary and rotating along the direction of grey arrow. Grey scale : trapping potential strength, non-dimensionalized by $f_0\sigma$, Colour-scale : radial component of particle velocities. Clusters are {\it{not}} bouncing at the boundary. (C,D)  Clusters are round in shape, rotating along the direction of grey arrow. Grey scale : trapping potential, non-dimensionalized by $f_0\sigma$, Colour-scale : radial component of particle velocities. Clusters are bouncing at the boundary that generates a force couple ---$(\mathcal{F}_r,\mathcal{F}_r^{\prime})$---(see D) within it. (D) Zoomed part of the lower half of (C) (denoted by the box with red broken line). The colour scale is same as (C). White arrow -- schematic trajectory of the cluster while bouncing at the boundary of the trap. Red arrow -- schematic direction of the radial velocities involved in the couple. Blue arrow : schematic velocity directions of the cluster along $\hat \theta$. Primed quantities : approaching towards the boundary before the bounce. Un-primed quantities : going away from the boundary after the bounce. Therefore the clusters roll around its centre of mass while rotating around the centre of the trap. (E) Typical trajectory of a single particle in a non-rolling but rotating cluster on XY plane. (F) Typical trajectory of a single particle in a rolling as well as rotating cluster on XY plane. (G) rolling frequency distribution in rolling cluster (black) and rolling frequency distribution in non-rolling cluster (red). (G(a)) Distribution of rolling frequency over the particles forming non-rolling cluster rotating along the direction of the grey arrows. Colour scale : rolling frequency. (G(b)) Distribution of rolling frequency over the particles forming rolling cluster rotating along the direction of the grey arrows. Colour scale : rolling frequency. Grey scale : trapping potential strength. non-dimensionalized by $f_0\sigma$ (G(c)) Zoomed in configuration from non-rolling cluster in (G(a)) where structural order is absent. (G(d)) Zoomed in configuration from rolling cluster in (G(b)) where structural order is present. (H) Standard deviation from the mean of the distribution of $P(v_r^i)$ for various steepness of the trap. Red open points : from simulations. Black solid line(Eq : $\sigma_{v_r}=A /\sqrt{S}+B, A=0.1598, B=-0.0753203$) : a fit from theory with via $A$ and $B$. }
\label{MainFig3}
\end{figure}
\end{center}

\twocolumngrid

It is apparent from {Fig{\ref{Phase1}}(A)} that there is a threshold value for the steepness ($S_c$) of the trap below which the structural order can emerge for Pe $\ge 20$. When $S > S_c$, apart from very low Pe, for all other Pe, particles accumulate at the boundary of the trap, spreads over it and flock together along the boundary. Thus we obtain polar clusters rotating along the circular boundary of the trap. The boundary of the trap, when steep (i.e. $S > S_c$), can exert repulsive force only on the interfacial particles i.e. the peripheral particles of the cluster which are closest to the trap boundary. The particles behind the interfacial particles are flocking together along the circular boundary under active force, inter particle collisions and thermal fluctuations. The direct influence of the repulsion from the boundary on them is very small. It is apparent from the Fig{\ref{MainFig2}(A,a)}, where we have shown the position of the particles of such clusters with their individual direction of self propulsion together with the values of the potential they are facing from the confining trap. When $S\geq S_c$, the particles follow the direction almost tangential to the trap. We find the interface between the cluster and the trap smooth Fig{\ref{MainFig2}(A,a)}.  

When the steepness of the trap approaches $S_c$, ripples start to appear at the interface, making it  rough Fig{\ref{MainFig2}(B,b)}. With lowering $S$, together with the interfacial particles, the other particles behind the interface start to feel the radially inward repulsion from the boundary. The repulsion decreases exponentially over a length scale $l$ as one goes radially inward, from the boundary of the trap. Therefore the particles feel spatially non-uniform repulsion from the boundary:  particles closer to (far from) the boundary face more (less) repulsion from it. This causes the particles which are close to the boundary at a certain instant of time, to go away from it in the next instants, where the repulsion is very small. But from there, due to self propulsion they again come close to the boundary and be repelled. Therefore due to the interplay between the self propulsion and non-uniform repulsion from the boundary of the trap, the particles are bouncing back and forth at the boundary. As the cluster is polar, even while bouncing, the particles are following each other efficiently. Therefore they bounce coherently at the boundary of the trap. Consequently ripples are formed at the interface of the polar cluster and the circular boundary of the trap. When $S$ is decreased further, more and more particles face non-uniform, radially inward repulsion from the trap boundary. Size of the ripples grows. Roughness of the interface enhances that finally leads to fragmentation of a cluster into smaller ones. This phenomena is clearly demonstrated in Fig{\ref{MainFig2}}. The fragmented clusters are also bouncing back and forth at the boundary due to non-uniform repulsion from the boundary and self propulsion of the particles. It is apparent from the oscillation of the position of their centre of mass over time, shown in {Fig{\ref{MainFig3}}(A, inset)}. In the same figure we have also plotted the position of  the centre of mass of the clusters that we have when the boundary of the trap is very steep i.e. $S > S_c$. The oscillation is negligibly small comparison to the former which implies the absence of bounce in the latter.   

When $S < S_c$, we have found that the fragmented polar clusters are repeatedly bouncing at the boundary of the trap. Therefore the radial component of the velocities ($v_r^i$) of the particles within the clusters can be positive as well as negative. The distribution of $v_r^i$ over the particles of the clusters are shown in {Fig{\ref{MainFig3}}(A, B, C)}. It shows that the particles of a cluster approaching the wall of the trap have positive radial velocities, particles at the wall have zero radial velocities and particles bouncing off the wall have negative radial velocities. Most of the particles in such cluster have non-zero $v_r^i$. Almost half of the particles of the cluster has $v_r^i\geq 0$ and the other half has $v_r^i < 0$. Therefore the distribution $P(v_r^i)$ (shown in {Fig{\ref{MainFig3}}(A)}) has the mean at zero and its standard deviation $\sigma_{v_r}$ is very large.  On the other hand, most of the particles of the clusters in the trap with steep boundary ($S > S_c$) have zero or very low radial velocities. The distribution $P(v_r^i)$ in this case, has zero mean and very small $\sigma_{v_r}$ in comparison to the former. 

 For the particles in a highly viscous suspensions, the velocity is proportional to the force where the proportionality constant is the mobility $\mu=\gamma^{-1}$ of the particles. Therefore the distribution of the radial component of the net force (say, $\mathcal{F}_r^i$) on the particles should be qualitatively same as radial velocity distribution. This implies that for almost half of the cluster that formed within a smooth ($S < S_c$) trap,  $\mathcal{F}_r^i \geq 0$. Particles of this half are approaching the boundary before the bounce. For the other half $\mathcal{F}_r^i < 0$. These particles are going away from the boundary after the bounce. Consequently a force couple is generated within the bouncing cluster {Fig{\ref{MainFig3}}(D)}. It generates a torque to roll around its own centre of mass. The force couple is absent or negligibly small within the clusters confined by the steep trap where $S > S_c$ because most of the particles have zero or very small radial velocities. Therefore no torque is generated within such clusters to roll.
      
One can find the evidence of the rolling of an individual cluster while plotting a typical trajectory of any particle within the cluster ({Fig{\ref{MainFig3}}(F)}). The trajectory on XY plane manifests superposition of two time periodic functions with different frequencies and phases. One of them represents its rotation (flocking) along the boundary which is about the origin of the trap. The other one represents its rolling about its own centre of mass. It is expected that this feature is absent for the cluster within a trap having $S > S_c$ ({Fig{\ref{MainFig3}}(E)}) as $\sigma_{v_r}$ is small in that case. In {Fig{\ref{MainFig3}}(G)}, for steep as well as smooth traps, we have calculated the frequency of rolling ($\omega_i$) with respect to centre of mass for the individual particles and show how they are distributed. From there it can be inferred that for steep traps, the variation of $\omega_i$ over the particles in the cluster (can be quantified by the standard deviation $\sigma_{\omega}$) is much smaller than its variation observed in the clusters within smooth traps. It implies that for most of the particles in a cluster within a steep trap, the rolling frequency is negligibly small. Only the interfacial particles have finite rolling frequency which is not enough for the whole cluster to roll. On the other hand, in case of the clusters within a smooth trap, almost all the particles have finite rolling frequency ($\sigma_{\omega}$ is large). Thus the whole cluster rolls around its centre of mass while rotating around the centre of the trap. Rolling makes the cluster round in shape.

One can also find how $\sigma_{v_r}$, for which the force couple and thereby the rolling of a cluster occurs, scales with the steepness of the trap. The energy involved in rolling of a cluster, $e_r\propto I\omega_c^2$ where $I$ is its moment of inertia and $\omega_c$ is its rolling frequency. The width of a cluster measured from its interface with the trap boundary depends on the steepness of the trap. It is proportional to $l =q^{-1}$ which implies, $I\propto m_cl^2$ where $m_c$ is the mass of the cluster. So $e_r\propto m_cl^2\omega_c^2$. Again, from previous discussion we know that the rolling of a cluster occurs due to a force couple. At moderately high Pe, when the clusters are formed, the magnitude of the active force $f_0$ exerted by individual particle, contributes to the force couple much more than other forces present in the system. In the scale of a single cluster, $f_0$ can be considered as the magnitude of active force per unit area. Therefore the total active force involved in a cluster is proportional to $f_0l^2$. Thus the energy involved in rolling can also be written as, $e_r\propto f_0l^3$. Equating the expressions for energies involved in rolling we obtain $\omega_c\propto f_0^{1/2}l^{1/2} = f_0^{1/2}q^{-1/2} \propto f_0^{1/2}S^{-1/2}$. From previous discussion, we know that the cluster rolls because of the large variation of radial velocities of the particles i.e. for large $\sigma_{v_r}$. Therefore, for simplicity if one assume $\sigma_{v_r}\propto \omega_c$, it provides $\sigma_{v_r} \propto S^{-1/2}$ for constant $f_0$.  We have tested the scaling relation between $\sigma_{v_r}$ and the steepness of the trap $S$ and the result is in {Fig{\ref{MainFig3}}(H)}.

From the discussions above, it is apparent that for $S < S_c$ and moderately high Pe, the rotating (along the circular boundary) polar clusters can roll around their centre of mass. We have shown that how the rolling frequency of a cluster scales with the steepness $S$ of the trap. The rolling is important. We find in our simulations that if a cluster rolls, then only structural order emerges within it. This is because the structural order to emerge with short range repulsive forces such as WCA, the particles should be close enough with each other making the cluster compact. In the current context, the circular boundary can compactify the rotating clusters only when they can roll. When $S > S_c$, though the clusters rotate along the boundary of the trap they cannot roll around its centre of mass. Therefore the particles within the cluster, which are away from the boundary of the trap, never face the confining potential directly. So they cannot be close enough to each other to exhibit the order. On the other hand, when $S < S_c$, for moderately high Pe, the clusters can rotate as well as roll along the boundary, facing the confinement through out its periphery over time. Thus it can be compactified by the circular boundary of the trap to manifest the order. \\

\section{Stability of order}

\onecolumngrid

\begin{center}
\begin{figure}[h]
\includegraphics[scale=0.50]{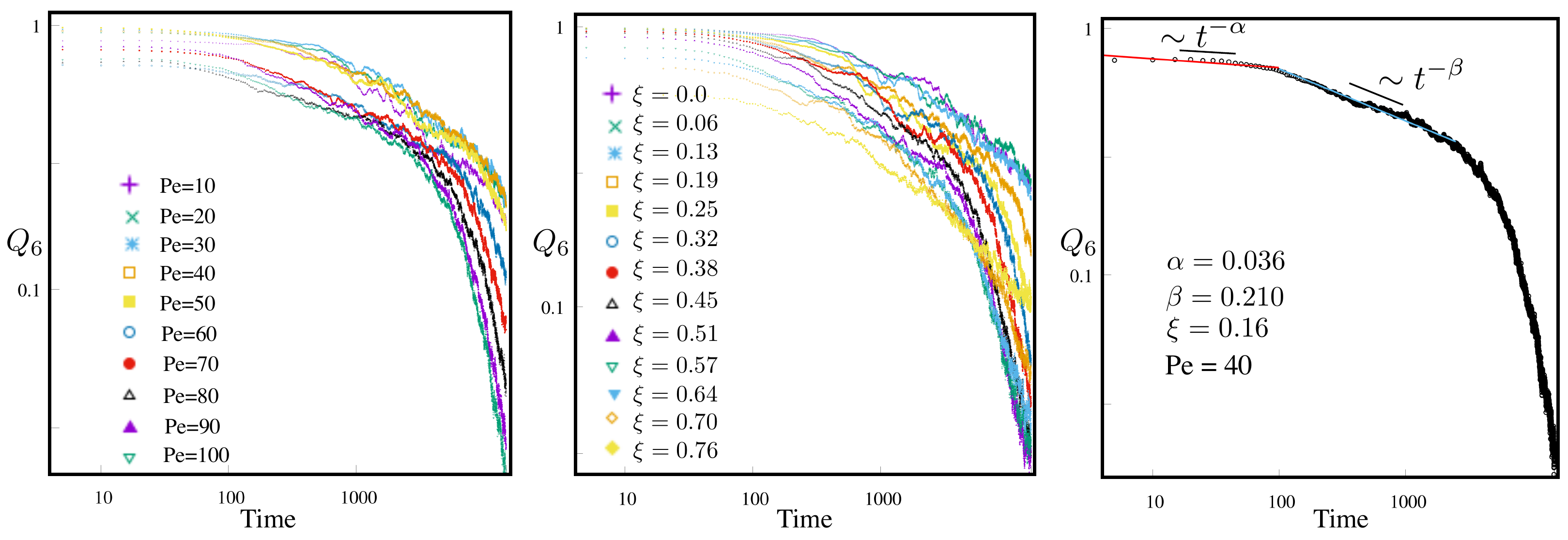}
\caption{{\bf{Decay of the structural order with time in log-scale}}($t$): (A) Decay of $Q_6$ with time for various Pe and $\xi=1/2\pi$. (B) Decay of $Q_6$ with time for various $\xi$ with Pe=40. (C) $Q_6$ with time steps for particular set of ($\xi $, Pe).}
\label{decay}
\end{figure}
\end{center} 

\twocolumngrid

In the sections before, we have discussed that tuning the steepness of the trap can give rise to the hexagonal order within the population of the self-propelling particles, in addition to the pre-established polar order. The relevant parameter space where the transition can occur, is identified and the kinetics of the onset of such transition is also detailed. Below we will discuss stability of the hexagonal order in absence of the trap. %a few important features of such transition. 

%{\bf(a)} 
Self-propelling but non-aligning particles can be trapped acoustically and when the concentration is high they crystallise to a close pack structure within the trap. It has been demonstrated that in such cases when the trap is switched off, the crystal {\it{explodes}} and homogeneous distribution of active particles are obtained quickly, within 12 seconds after switching of the trap \cite{takatori2016acoustic}. 
Here the hexagonal order is the outcome of the interplay between the self-propulsion and velocity alignment interactions of the particles together with the steepness of the trap. The ordering takes place at low average packing fraction. Such an order is unstable if any one of the trio is absent. One can investigate that how the order being unstable decays over time, when the trap is suddenly switched off. We consider that initially we prepare the system such that the particles are in a hexagonally ordered, polar cluster.  Then suddenly the trap is switched off and we calculate $Q_6$ over time. In {Fig{\ref{decay}}}  we plot the time evolution of $Q_6$ after switching off the trap. We find that $Q_6$ decays with power law. Interestingly we find that there are two distinct regimes of the decay where it follows two different powers. The regimes crosses over each other. It indicates that possibly multiple physically different processes are responsible for the decay. More systematic study is needed along this line to identify the exponents and the physical processes, which is not included in the present article.

%{\bf{(b)}} We have shown that the cluster of the active particles  within a steep trap (i.e. $S > S_c$) are strip-like and it spreads over the circular boundary. Such clusters are polar but they have no structural order. There are two length scales of such clusters --- its length along the circular boundary which goes as $r_0$ and its width, which is proportional to $l \sim S^{-1}$ i.e. the width of the cluster depends on the steepness of the trap. 
%When $S < S_c$, the polar cluster becomes round in shape and structurally ordered as well. It has only one length scale which is proportional to $l \sim S^{-1}$ . We find from the simulations that the degree of the structural order within the clusters depends on $l$ and therefore on $S$ as well and the number of such clusters depend on the ratio $l/r_0$. In particular, keeping the packing fraction $\phi$ same, if $r_0$ is altered (say, to investigate the finite size effect ), in order to obtain same number of the clusters with same size, $l$ should be altered keeping $l/r_0 << 1$, constant. \\           

%\subsection{Continuum Vicsek and discrete Vicsek}
%\subsection{Active Quincke effect ?}
%\subsection{Cluster size and number - $l/r_0$ }

\section{Conclusion}

The model and the results discussed here is amenable to the experiments.  For example, one may consider induced-charge, electrophoretic, self-propelled Janus colloidal particles, as described in \cite{van2019interrupted}  within a confinement providing electrostatic repulsion to the particles. The repulsion decays exponentially with the distance between the particle and the confining boundary. The decay involves a length scale (Debye length) that typically depends on the solution, the size of the particles and the zeta potential of the surface of the colloidal particle and the boundary \cite{rashidi2017motion}. Thus the decay length may be tuned to control the steepness of the confining boundary, which is essential for experimental realisation of the system discussed here. 

We have considered a circular (thereby, spherically symmetric) trap with tuneable steepness, which can confine self-propelling, self-aligning, soft discs at low packing fraction. The particles being active, accumulate at the boundary of the steep trap. They form strip-like polar cluster that spreads and rotate along the boundary. The polar order is stable because their rotational fluctuation is quite small compared to self propulsion and velocity aligning interaction. Tuning the steepness of the trap we have shown that as the steepness goes below than a threshold value, the cluster becomes round in shape and hexagonal order emerges within it, keeping the polar order intact. The transition occurs because, as the steepness of the trap is reduced, the polar, strip-like cluster is fragmented  into smaller clusters which not only rotate but also roll along the boundary of the trap. It is the rolling for which the trap can push the particles inside a cluster, from every direction, as it rotates along the trap boundary. Due to this, in addition to the polar order, the cluster becomes compact enough to exhibit hexagonal order in the presence of short-ranged, isotropic, inter-particle repulsion. The hexagonal order is unstable in the absence of the boundary. When the trap is switched off, the order decays over time with power law.

\section{ACKNOWLEDGMENTS}
AS thanks University Grants Commission Faculty Recharge Program (UGCFRP), India.

\bibliography{frozen}

\end{document}